\documentclass{arXiv}

\journalname{JGR-Space Physics}

\begin{document}

%
%

\title{\textcolor{black}{Global three-dimensional simulation of Earth's dayside reconnection using a two-way coupled magnetohydrodynamics with embedded particle-in-cell model: initial results}}

%
%

 \authors{Yuxi Chen\affil{1},
 {G\'abor T\'oth} \affil{1}, 
  Paul Cassak\affil{2}, 
  Xianzhe Jia\affil{1},
  Tamas I. Gombosi\affil{1},                                              
  James A. Slavin\affil{1},                                                
  Stefano Markidis\affil{3},
  Ivy Bo Peng\affil{3},
  Vania K. Jordanova\affil{4}
  }    

\affiliation{1}{Center for Space Environment Modeling, University of Michigan,                                                                
Ann Arbor, MI 48109, USA}

\affiliation{2}{                                                                                                         
Department of Physics and Astronomy, West Virginia University, Morgantown, WV 26506, USA.}                                     
                                                                                                                         
\affiliation{3}{                                                                                                         
KTH, Stockholm, Sweden.} 

\affiliation{4}{
Los Alamos National Laboratory, Los Alamos, NM 87545, USA}


\correspondingauthor{Yuxi Chen}{yuxichen@umich.edu}


\begin{keypoints}
\item A one-hour long global simulation of Earth's magnetosphere with kinetic modeling of the dayside reconnection
\item Crater FTE is found at the early stage of a flux rope formation
\item Kinetic phenomena are found from the global simulation
\end{keypoints}

%
%

\begin{abstract}
We perform a three-dimensional (3D) global simulation of Earth's magnetosphere with kinetic reconnection
physics to study the flux transfer events (FTEs) and dayside magnetic reconnection
with the recently developed magnetohydrodynamics with embedded particle-in-cell model (MHD-EPIC). 
During the one-hour long simulation, the FTEs are generated quasi-periodically near 
the subsolar point and move toward the poles. We find the magnetic field signature
of FTEs at their early formation stage is similar to a `crater FTE', 
which is characterized by a magnetic field strength dip at the FTE center. After the FTE 
core field grows to a significant value, it becomes an FTE with typical flux rope structure. 
When an FTE moves across the cusp, reconnection between the FTE field lines and the cusp 
field lines can dissipate the FTE. The kinetic features are also captured 
by our model. A crescent electron phase space distribution is found near the 
reconnection site. A similar distribution is found for ions at the location where the Larmor
electric field appears. The lower hybrid drift instability (LHDI) along the 
current sheet direction also arises at the interface of magnetosheath and magnetosphere
plasma. The LHDI electric field is about 8 mV/m and its dominant wavelength relative to the electron gyroradius agrees reasonably with MMS observations.


\end{abstract}

%
%

%



%
%
%

\section{Introduction}
Magnetic reconnection between the interplanetary magnetic field (IMF) 
and Earth's intrinsic dipole field is regarded as the most important 
mechanism for mass and energy transfer from the solar wind to the
magnetosphere. Flux transfer events (FTEs) are widely considered 
as a phenomenon related to dayside non-steady reconnection \citep{Russell:1978}. 
An FTE is a bundle of reconnected magnetic fluxtubes created at the magnetopause 
and moving anti-sunward along the magnetopause. Such events are 
characterized by a bipolar variation of the magnetopause normal magnetic field $B_N$, 
and are usually associated with an enhancement of core field, the 
magnetic field component along the axial direction of the FTE. 
An FTE exhibits a flux-rope structure in three-dimensional space.
It has been observed that the plasma 
inside an FTE is usually a mixture of magnetospheric and 
magnetosheath plasma \citep{Daly:1981}, indicating that FTEs are 
generated by magnetic reconnection process. The diameter of an FTE can
 vary from several ion inertial lengths \citep{Eastwood:2016} 
(a few hundred kilometers) to several Earth radii \citep{Rijnbeek:1984,Hasegawa:2006}. 
In the dawn-dusk direction along the magnetopause, FTEs
can extend over a long distance \citep{Fear:2008}. FTEs frequently occur as a quasi-periodic process, and \citet{Rijnbeek:1984} reported that the FTEs were observed 
about every 8 minutes \textcolor{black}{during periods of 
southward magnetosheath magnetic field}.\\

FTEs have been studied with various global numerical models. 
Compared to local simulations, a global model can offer more 
realistic plasma and magnetic field context. \citet{Fedder:2002} 
used a global ideal MHD model to study the generation of FTEs. 
The typical magnetic field signature is captured by their model, 
and their simulation suggests that the FTEs are formed by non-steady 
reconnection along the separator at the magnetopause. 
\citet{Raeder:2006} performed a high resolution ideal MHD simulation 
with the OpenGGCM model. FTEs formed by multiple X line reconnection \citep{Lee:1985} with 
a tilted dipole field in this study. \citet{Dorelli:2009} revisited 
the FTE generation mechanism with resistive MHD using the OpenGGCM model, 
and the authors argue that
the FTEs are generated by flow vortices and the formation of new X lines 
is the consequence, rather than the cause of FTE formation. 
\citet{Sibeck:2008} studied crater FTEs with the BATS-R-US MHD model. 
All these global simulations are based on ideal or resistive MHD codes, 
and the generation of FTEs relies either on ad hoc resistivity \citep{Dorelli:2009} or 
numerical resistivity \citep{Fedder:2002, Raeder:2006}. Recently, 
a 2D-3V global magnetospheric hybrid-Vlasov simulation was performed to 
study magnetopause reconnection and FTEs by \citet{Hoilijoki:2017}.\\


Typical FTEs are associated with an enhancement of the field strength at the center of a flux rope. On the other hand, the so-called crater FTEs show more complicated structure: the center field is surrounded by two `trenches' and the field strength usually show a dip just at the center \citep{Labelle:1987,Owen:2008}. Typical FTEs are more frequently observed than crater FTEs \citep{Zhang:2010}. The generation mechanism of crater FTEs has been explored with both numerical simulations \citep{Sibeck:2008} and analytic models \citep{Zhang:2010}. \citet{Zhang:2010} proposed that crater FTEs are the initial stage of typical FTEs based on hundreds of events selected from THEMIS observations. The structure of the core field can be even more complicated, for example, \citet{Eriksson:2016} found a tripolar core field flux rope at the magnetopause.\\

It is widely accepted that the formation of FTEs is related to the dayside magnetopause reconnection, which is a kinetic process for collisionless plasma. Therefore it is important to include proper kinetic effects into the numerical model in order to produce FTEs in a physical way. The MHD with embedded PIC (MHD-EPIC) model developed by \citet{Daldorff:2014} makes it feasible to use a kinetic model to study reconnection and FTEs with realistic magnetospheric configuration for the first time. Because of the small kinetic scales inside the magnetosheath, for example, the ion inertial length $d_i$ is about 60km $\sim 1/100\,R_E$, we have to artificially increase the kinetic scales in the present 3D global simulation. As shown by our companion paper \citep{Toth:2017}, this scaling has on significant effect on the large scale structures, while the kinetic phenomena occur at linearly increased scale. Since the kinetic scale physics is included in our global model, the reconnection related kinetic phenomena, like the crescent shape electron phase space distribution, the Larmor electric field and the lower hybrid drift instability (LHDI), are all captured by the model. The crescent distribution was first found by \citet{Hesse:2014} from 2D local simulation, then observed by the Magnetospheric Multiscale (MMS) mission recently \citep{Burch:2016}. It is formed by the magnetosheath electrons reaching the stagnation point and accelerated by the Hall electric field \citep{Bessho:2016,Shay:2016}. This special distribution has been proposed as an indicator of the magnetic reconnection location \citep{Hesse:2014}. The Larmor electric field is potentially another signature that can help to identify the location of reconnection site \citep{Malakit:2013}. It is on the magnetosphere side, normal to the current  sheet and pointing away from the X line. The lower hybrid drift instability (LHDI) develops along the current direction \citep{Daughton:2003,Roytershteyn:2012}, and it has been observed recently by MMS satellites \citep{Graham:2016}. LHDI was considered as a potential source to create anomalous resistivity for reconnection \citep{Huba:1977}, but previous research \citep{Mozer:2011} has suggested the related resistivity may be not large enough. However a recent 3D simulation showed LHDI may still play an important role near the diffusion region because of the presence of turbulence \citep{Price:2016} .\\

In the following sections we will describe the MHD-EPIC model, the simulation setup, and then discuss the simulation results. 

\section{Model description}
The MHD-EPIC model has been successfully applied to investigate the interaction between the Jovian wind and Ganymede's magnetosphere, where the ion inertial length is large compared to the size of its magnetosphere \citep{Toth:2016}. In this paper, the same model is applied to study Earth's magnetosphere, which is more challenging because of the small kinetic scale. The MHD-EPIC model two-way couples the BATS-R-US \citep{Powell:1999, Toth:2008} MHD code and the implicit particle-in-cell code iPIC3D \citep{Markidis:2010} through the Space Weather Modeling Framework (SWMF) \citep{Toth:2005swmf, Toth:2012swmf}. A general description of the these models and the simulation setup is provided in this session. 

\subsection{Global MHD model: BATS-R-US}
In order to make the MHD model as complete as possible, the Hall term and the electron pressure gradient term are included in the generalized Ohm's law, and a separate electron pressure equation is solved. The generalized Ohm's law we use is:
\begin{linenomath*}
\begin{equation} \label{eq:Ohm}
\mathbf{E} = -\mathbf{u}\times \mathbf{B} + \frac{\mathbf{J}\times \mathbf{B}}{q_e n_e}-\frac{\nabla p_e}{q_e n_e}
\end{equation}
\end{linenomath*}
where $q_e$, $n_e$ and $p_e$ are the charge per electron, electron number density and electron pressure, respectively. The electron pressure is obtained from:
\begin{linenomath*}
\begin{equation} \label{eq:pe}
\frac{\partial p_e}{\partial t} + \nabla \cdot (p_e \mathbf{u}_e) = (\gamma -1) (-p_e \nabla \cdot \mathbf{u}_e )
\end{equation}
\end{linenomath*}
where $\gamma = 5/3$ is the adiabatic index, and $\mathbf{u_e} = \mathbf{u} - \mathbf{J}/(q_en_e)$ is the electron velocity.\\

 From the numerical prospective, it is not trivial to incorporate the Hall term into the MHD equations. The Hall MHD equations support the whistler mode wave, which is dispersive and the characteristic speed is inversely proportional to the wave length. Since the shortest wave length that can be resolved in a numerical system is twice the cell size, the fastest whistler wave speed is proportional to $1/\Delta x$. For an explicit time integration scheme, the time step is limited by the CFL condition, which leads to a time step approximately proportional to  $1/(\Delta x)^2$ for Hall MHD.  In order to use a reasonably large time step, a semi-implicit time discretization is employed \citep{Toth:2012swmf}. The semi-implicit scheme treats the stiff terms, which is the Hall term here, and other terms separately. Excluding the Hall term, the rest of the equations are updated with an explicit scheme, and the time step is only limited by the fast magnetosonic wave speed. The Hall term is handled by an implicit solver after the explicit update has been done. \\
 
\begin{figure}[h]
\includegraphics[height=3.5in]{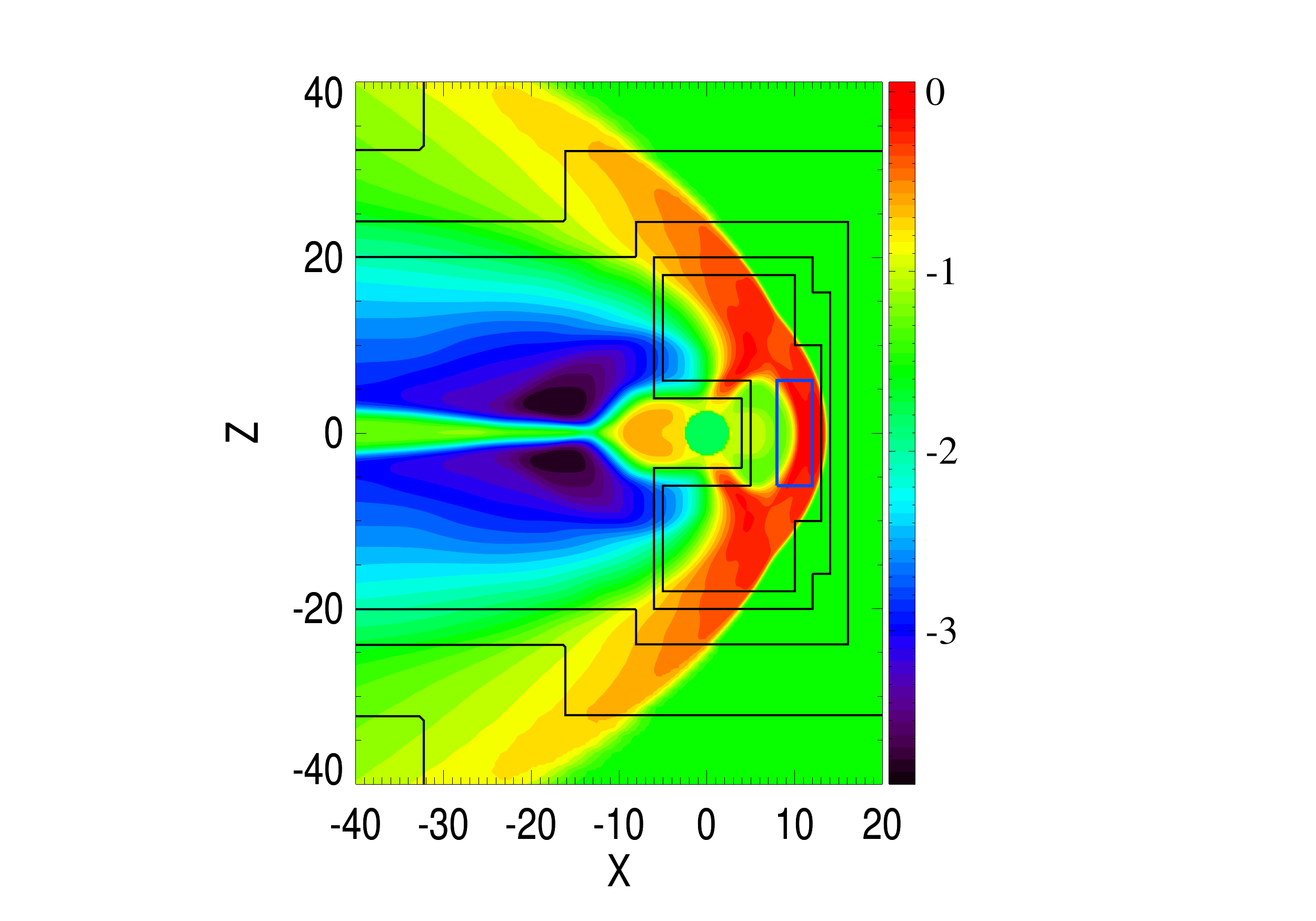}
\caption{Part of the meridional plane with the adaptive MHD grid and the PIC region. The color represents the plasma pressure on a logarithmic scale. The black lines represent the refinement level, where the cell size changes. The resolution of the finest level around the dayside magnetopause is $1/16\,R_E$, and the refinement ratio between two nearby levels is 2. The blue box ($8\,R_E<x<12\,R_E$, $-6\,R_E<z<6\,R_E$) is the edge of the PIC region covered by iPIC3D, and it extends from $-6\,R_E$ to $6\,R_E$ in the y direction.}
\label{fig:grid}
\end{figure}

The typical solar wind condition at 1AU with purely southward IMF is used as the boundary condition to drive the magnetosphere: $\mathbf{B}=(0,0,-5)$ nT, mass density $\rho = 5\,\mathrm{amu/cm^3}$, ion pressure $p_i=3.45 \times 10^{-3}$ nPa, and solar wind velocity $\mathbf{u}=(-400,0,0)$ km/s. Electron pressure $p_e=8p_i=2.76 \times 10^{-2}$ nPa is used, so that after crossing the shock, where the ions are heated by converting bulk into thermal energy while the electron thermal energy changes adiabatically, the ion-electron pressure ratio is about $p_i/p_e \sim 2.5$.  \citet{Wang:2012} shows that the temperature ratio $T_i/T_e$ in the solar wind varies from $0.1\sim 2$, and the ratio is about $4\sim 12$ inside the magnetosheath. The $T_i/T_e$ ratio, which is the same as $p_i/p_e$, used in the simulation is close to but slightly smaller than the typical observed ratio. We use $T_i/T_e=1/8$, because our numerical experiments show that the electrons can be numerically heated by PIC if colder solar wind electrons are used as boundary condition. A magnetic dipole with 30116 nT field strength at the magnetic equator is used. Its magnetic axis is aligned with the z axis. The total magnetic field $\mathbf{B}$ is split into the intrinsic dipole field $\mathbf{B_0}$ and the deviation $\mathbf{B_1}$. A three-dimensional block-adaptive Cartesian grid is used to cover the whole magnetosphere: $-224\,R_E<x<32\,R_E$, $-128\,R_E<y<128\,R_E$ and $-128\,R_E<z<128\,R_E$. Since we focus on the dayside dynamics in this paper, the mesh along the dayside magnetopause is refined to high resolution with $\Delta x = 1/16\,R_E$ (see Figure~\ref{fig:grid}). 59 million cells are used in total. At the inner boundary $r=2.5\,R_E$, the density is fixed as $28\, \mathrm{amu/cm^3}$, the pressure and the magnetic field $\mathbf{B_1}$ have zero gradient, the radial velocity is zero, while the tangential velocity is calculated from the ionosphere electrodynamics model developed by \citet{Ridley:2004}.\\

\subsection{Implicit particle-in-cell model: iPIC3D}\label{sec:iPIC3D}
The semi-implicit particle-in-cell code iPIC3D was developed by \citet{Markidis:2010}. The advantage of iPIC3D over  explicit particle-in-cell codes is that iPIC3D is linearly unconditionally stable, so that iPIC3D can handle larger time step and larger cell size than explicit PIC. Compared to the explicit PIC method, the cell size of iPIC3D is chosen based on the scale of interest instead of the Debye length, and the time step of iPIC3D is not limited by the plasma frequency or the speed of light, but the accuracy condition, which requires ${v}_{rms} \Delta t/\Delta x<1$ on all grid nodes for all species, where ${v}_{rms}$ is the root mean square of macro-particle velocities. In order to make the simulation as efficient as possible while keeping the accuracy condition satisfied, we use an adaptive time step:
\begin{linenomath*}
\begin{equation} \label{eq:cfl}
\Delta t = c_0 \cdot\min ( \Delta x /v_{rms},\, \Delta y /v_{rms}, \,\Delta z /v_{rms})
\end{equation}
\end{linenomath*}
calculated for each grid nodes and the minimum is taken over the whole PIC mesh. The root mean square velocity $v_{rms}$ is similar to the thermal velocity but contains the effect of bulk velocity. $c_0$ is a coefficient that should be smaller than 1. $c_0=0.4$ is used for the simulation in this paper.\\

Since the focus of this paper is the dayside magnetopause reconnection, the embedded PIC box is placed near the sub-solar magnetopause, where reconnection happens under purely southward IMF. In the GSM coordinates, the region inside $8\,R_E<x<12\,R_E$ and $-6\,R_E<y,z<6\,R_E$ (see Figure~\ref{fig:grid}) is solved by iPIC3D. The PIC region covers the magnetopause and it is just inside the bow shock. Since the size of the ion diffusion region is the same order as the ion inertial length, such kinetic scale should be resolved in order to capture reconnection kinetic physics. However, the ion inertial length $d_i=c/\omega_{pi}$ is about 60km $\sim 1/100\,R_E$ for a typical magnetosheath density of $20\,\mathrm{amu/cm^3}$. The length is so small that it is extremely difficult to resolve even for a 3D global MHD model, not to mention the PIC code. Scaling up the kinetic length helps to reduce computational resources. In the present simulation, all the fluid values, including density, pressure, velocity, IMF and dipole field strength, hence the derived values like the sound speed, Alfven velocity and plasma beta, are realistic so that the global structure of the magnetosphere is comparable to the real situation. On the other hand, the ion inertial length is scaled up 16 times to about $1/6\,R_E$ in the magnetosheath by artificially increasing ion mass per charge by a scaling factor of 16. Since all the quantities are normalized in the numerical model, there are several ways to understand or interpret the scaling. One way is treating the scaling as  changing the charge of ions and electrons. Compared with the original system, we reduce the charge by a factor of 16 while all the other basic physical quantities, like mass per ion, number density, and temperature remain realistic. From the perspective of ideal magnetohydrodynamics, the scaled system is exactly equivalent to the original one. For a particle-in-cell code, the reduction of charge per ion reduces the electromagnetic force on an ion and therefore increases the gyroradius and gyroperiod by a factor of 16. But the gyroradius and the gyroperiod are still several orders smaller than the global spatial and temporal scale, for example the distance from Earth to the magnetopause and the time for the plasma moving from the subsolar point to the cusp, respectively. How the scaling changes the structure of reconnection is discussed in details in our companion paper by \citet{Toth:2017}. We also apply a reduced ion-electron mass ratio $m_i/m_e=100$, which is sufficiently large to separate the electron and ion scales. We choose $\Delta x=1/32\,R_E$ as the PIC grid resolution so that $\Delta x/d_i \sim 5$ and $\Delta x/d_e \sim 0.5$. This resolution keeps a balance between the computational cost and the requirement of resolving kinetic scales. 216 particles per cell per species are used and there are about 9 billion particles in total inside the domain initially. Our numerical experiments suggest smoothing the electric field $\mathbf{E}$ and the current density $\mathbf{j}$ can help to suppress the numerical noise \citep{Toth:2017}.\\

The typical magnetic field strength in the magnetosheath is about 30 nT, and the corresponding ion gyro-frequency is $\Omega_{ci}=0.0286$Hz and  $\Omega_{ce}=2.86$Hz with scaled charge-mass ratio. As mentioned above, the time step of iPIC3D is determined by the accuracy condition (Eq. \ref{eq:cfl}). From the simulation, we find the maximum thermal speed of electrons inside the PIC domain is about 2500km/s, which leads to a time step of $\Delta t \sim 0.03s \sim 10^{-3}\Omega_{ci}^{-1} \sim 0.1\Omega_{ce}^{-1}$ with cell size $\Delta x = 1/32\,R_E$. Therefore, the time step is small enough to resolve the gyro-motion of both electrons and ions.

\subsection{Coupling between BATS-R-US and iPIC3D}
BATS-R-US and iPIC3D are coupled through the Space Weather Modeling Framework (SWMF). These two models are compiled together to generate a single executable file. Both models can run simultaneously on specified processors and the information exchange is paralleled and handled by the Message Passing Interface (MPI). The details of the two-way coupling has been described by \citet{Daldorff:2014}.\\

In the simulation presented in this paper, we run the Hall MHD code first with the local time stepping scheme to reach a steady state. Then BATS-R-US sends the information, including density, velocity, pressure and magnetic field, to iPIC3D. iPIC3D initializes the electric field based on the Ohm's law. The Maxwellian distributed particles are generated according to the fluid information so that iPIC3D and BATS-R-US have consistent density, velocity and pressure at the same position. After the PIC initialization, the MHD and PIC models update independently with their own time steps. The coupling frequency between these two models can be set to a value that is independent of the MHD or PIC time step. During the coupling, iPIC3D calculates moments of the particle distribution function, such as the density, velocity and pressure, and overwrites the MHD cells overlapped with PIC region. In return the MHD model provides electromagnetic field as well as particle boundary conditions for iPIC3D. For the particle boundary, iPIC3D removes the particles in the boundary cells, and re-generates new particles based on the fluid variables obtained from MHD.  Between the two coupling time points, iPIC3D uses the latest information obtained from BATS-R-US as boundary condition during each iteration. In the simulation presented here, the time step for BATS-R-US and iPIC3D are around $\Delta t_{MHD}=0.015$ s and $\Delta t_{PIC}=0.032$ s, respectively. The coupling time interval is set to a small value $\Delta t_{couple}=0.005$ s so that MHD and PIC are coupled every time step. We note that the time step of PIC is even larger than the MHD, because the MHD time step is limited near the magnetic poles due to the high Alfven speed, while these regions are outside the PIC domain. \\

We used to generate particles in only one ghost cell layer \citep{Daldorff:2014} as particle boundary condition. Our numerical experiments suggest using more layers (5 layers specifically in this paper) as the particle boundary, while the electromagnetic field boundary is still only enforced at the outermost layer, is helpful to smoothly transit from PIC to MHD. The MHD cells overlapped with the PIC particle boundary are not overwritten by PIC. Similar technique has been used to implement open boundary condition for stand-alone PIC simulations \citep{Peng:2015}.\\

We run the simulation on 6400 processors for 170 hours to model one hour simulation time on Blue Water supercomputer \citep{Bode:2012}. iPIC3D and BATS-R-US use about $80\%$  and $15\%$ of the simulation time, respectively. The coupling and other overheads use the remaining $5\%$.\\

\subsection{Energy conservation}
\begin{figure}[htb]
\centering
    \includegraphics[width=0.6\textwidth, trim=0cm 0cm 0cm 0cm,clip,angle=0]{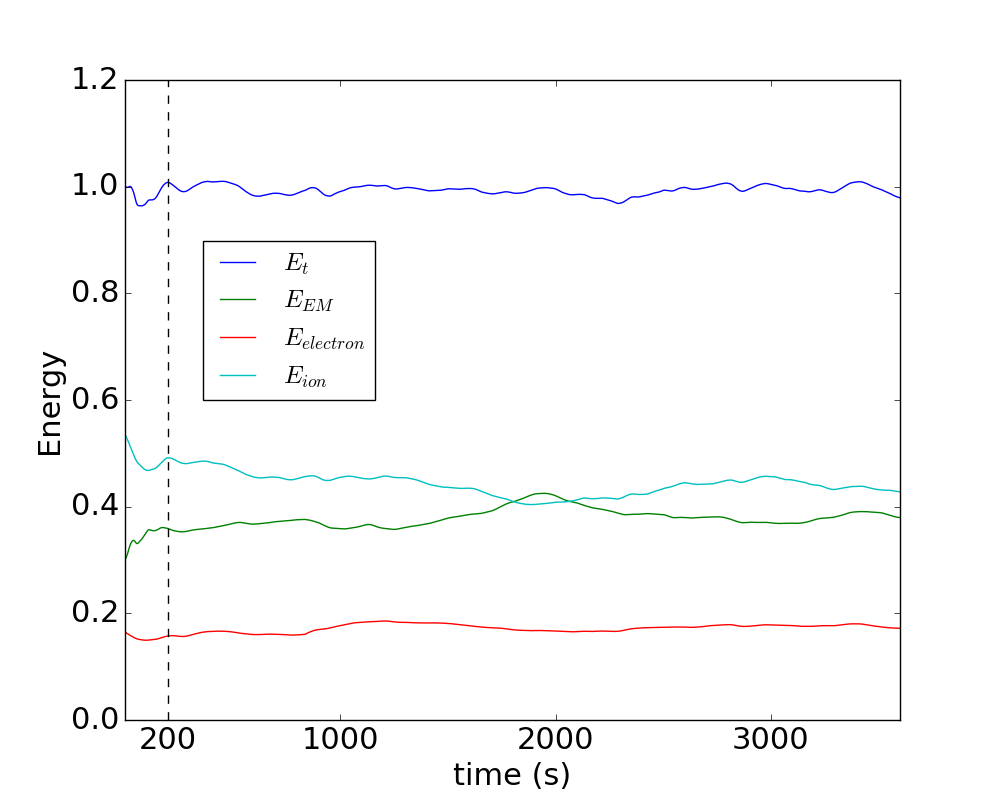} 
  \caption{The normalized the total energy $E_t$, electric field and magnetic field energy $E_{EM}$, ion energy $E_{ion}$ and electron energy $E_{electron}$. They are normalized by the initial total energy.}
  \label{fig:energy}
\end{figure}

Even though the PIC region is not a closed system, therefore mass and energy flow into and out of the region, it is still important to check the energy variation during the simulation to make sure the PIC model does not suffer from numerical heating or cooling. The normalized energy changes are shown in Figure~\ref{fig:energy}. Throughout the simulation, the total energy $E_t$ variation is less than $3\%$. The small variation suggests that the numerical heating or cooling are insignificant. The initial condition for iPIC3D is under MHD equilibrium, but not necessarily under Vlasov equilibrium. The electromagnetic field energy $E_{EM}$ and kinetic energy of each species normalized by the initial total energy are also shown in Figure~\ref{fig:energy}. During the first several minutes, energy is transferred from the particles to the electromagnetic field. After 200s, the ion and electron energy decreases about $5\%$, while the electromagnetic field energy increases from 0.3 to about 0.36. This is the transition from the MHD steady state to a PIC preferred solution. Further change of these energies are gradual and small. $E_{EM}$ is mainly magnetic field energy, which is about 3 orders larger than electric field energy.\\

\section{Results}
\subsection{Overview}
\begin{figure}[htb]
\centering
 \includegraphics[width=1.0\textwidth, trim=0cm 0cm 0cm 0cm,clip,angle=0]{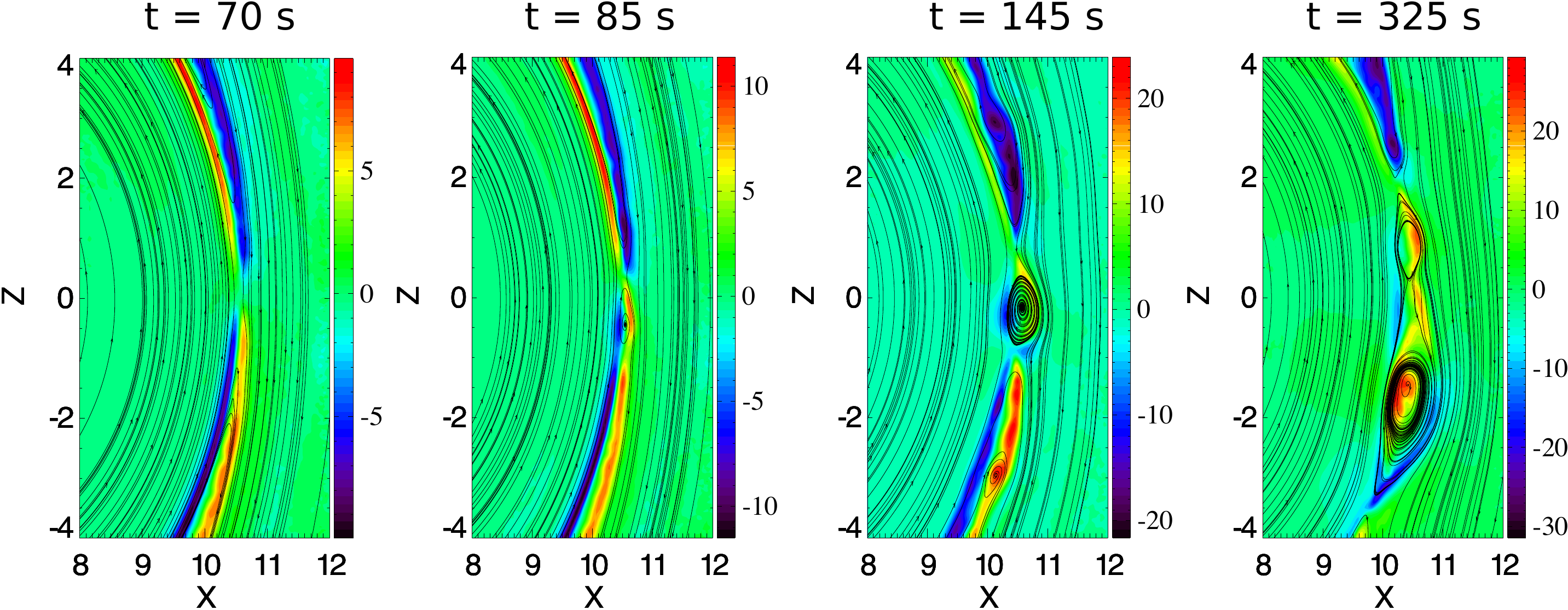} 
  \caption{A series of snapshots showing $B_y$ strength and the projected magnetic field lines in the meridional plane inside the PIC region. The color bar is different in each plot.}
  \label{fig:over-view}
\end{figure}

The iPIC3D code is initialized from a steady Hall MHD state, which is shown in Figure~\ref{fig:grid}. The steady state is obtained from the Hall MHD run by using a local time stepping scheme, and a reconnection X line already exists near the equatorial plane along the dayside magnetopause. Since the local time stepping scheme is diffusive in this case, the reconnection signature near the X line is weak, for example, the Hall magnetic field strength is only about 1 nT. The PIC code inherits the magnetic field topology and starts evolving based on Maxwell's equations and the motion of the macro-particles. An overview of the evolution of the dayside magnetopause is shown in Figure~\ref{fig:over-view}, which contains the Hall magnetic field $B_y$ and the field lines at the meridional plane inside the PIC box. At $t=70$ s, $B_y$ has already increased to about 8 nT. The Hall field extends far away from the X line with roughly the same field strength for each branch. 15s later, south of the existing reconnection point, another X line starts to form at around $x=10.2\,R_E$ and $z=-1\,R_E$. At $t=145$ s, both X lines can be seen clearly, and a flux rope like structure forms between the two X lines. The top X line has moved to about $z=0.5$. The bottom X line is almost steady so far, but it will move southward later. At $t=325$ s, the top and bottom X lines reach about $z=1.8$ and $z=-3.5$, respectively, and the center of the flux rope is moving southward with the bottom X line. Since the flux rope is moving away from the top X line, the current sheet between them  becomes unstable and a secondary flux rope is generated (rightmost panel of Figure~\ref{fig:over-view}).  During the one hour simulation, flux ropes form near the subsolar point and move toward poles quasi-periodically. More details about reconnection and flux ropes, both macroscopic and microscopic scales, are discussed in the following sub-sections. \\

\subsection{Evolution of FETs}
\begin{figure}[htb]
\centering
    \includegraphics[width=1.1\textwidth, trim=0cm 0cm 0cm 0cm,clip,angle=0]{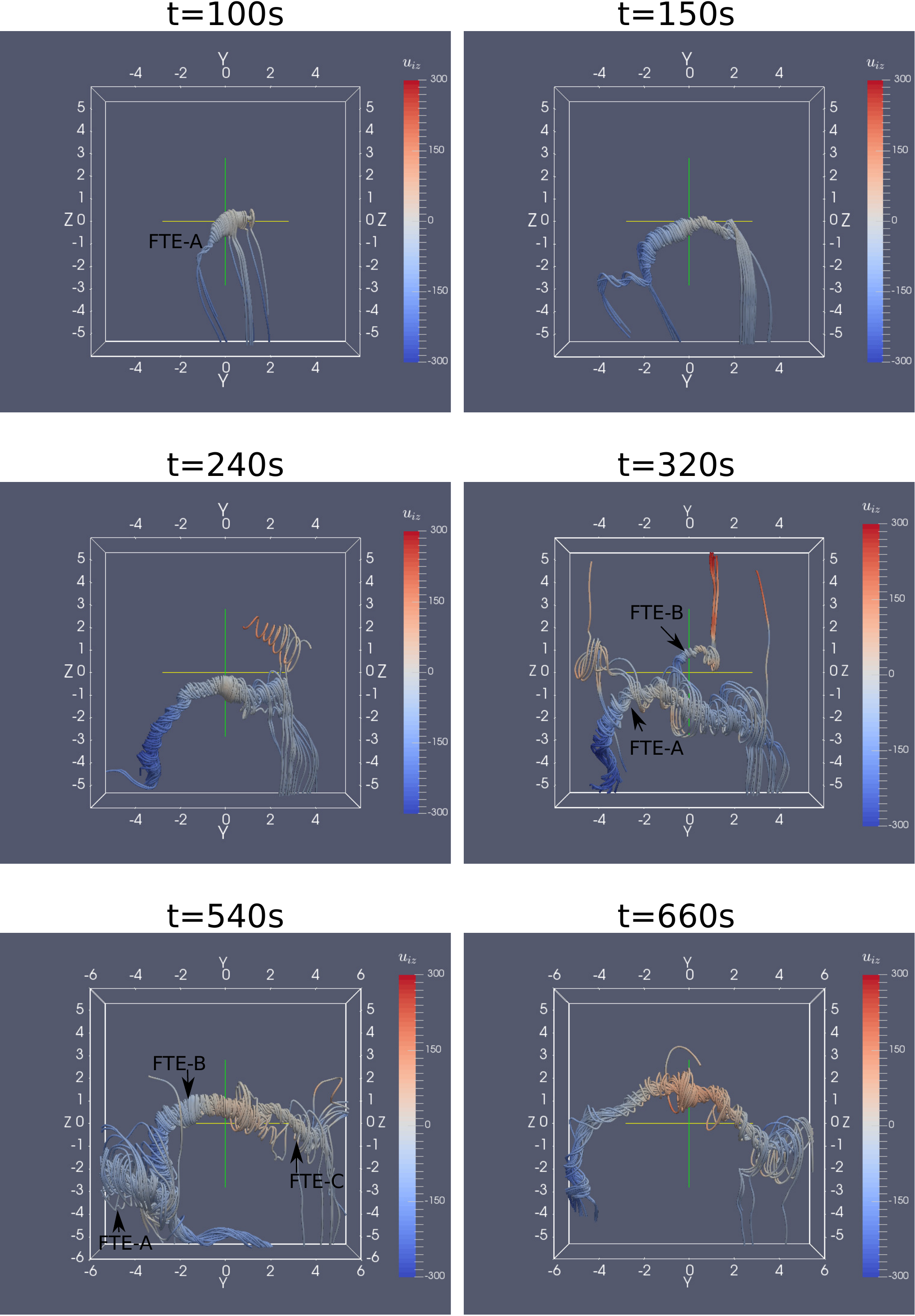}
  \caption{The evolution of FTEs. Viewed from the Sun, a series of snapshots with magnetic field lines colored by ion velocity $u_{iz}[\mathrm{km/s}]$ are shown.}
  \label{fig:fr-3d}
\end{figure}

The meridional cut of the first two FTEs formed in the simulation are already shown in Figure~\ref{fig:over-view}. When we go beyond the 2D view, more complicated but completed structures arise. The flux ropes colored with the ion velocity z component $u_{iz}$ at different times are shown in Figure~\ref{fig:fr-3d}. At $t=100\, \mathrm{s}$, a short flux rope appears near the subsolar point. It is labeled as FTE-A. This flux rope extends from $y\sim -1\,R_E$ to $y \sim 1\,R_E$ in the dawn-dusk direction. It suggests that next to the primary X line near $z=0$, another X line starts to form south of the subsolar point. We have checked a series of 2D $x-z$ plane cuts, and found that the signature of reconnection, like the ion jets, at the second X line is clear at $y=0$, but appears very weak far away from the Sun-Earth line, for example at $y=0.78\,R_E$ or $y=-0.78\,R_E$. At $t=150$ s, the flux rope has extended significantly in both dawn and dusk directions. Along the flux rope, the ion velocity varies. Close to the dusk side (positive y), the reconnection at the second X line produces fast northward ion jet flow to slow down the southward flow from the primary X line, so that the flux rope moves relatively slowly. Close to the dawn side (negative y), the reconnection at the second X line is not strong enough to offset the southward flow ejected from the primary X line. The varying ion velocity leads to an inclined flux rope. At $t=240$ s, the flux rope is even more tilted because of the varying ambient ion jet velocity. A new small flux rope, FTE-B in Figure~\ref{fig:fr-3d}, is generated at $t=320$ s above FTE-A. FTE-A bifurcates at $y\sim -2.5$ and the new branch extends along the dawn-northward direction. FTE-A keeps moving southward while FTE-B is growing. At $t=540$ s, a large portion of FTE-A, except for the dawn part, already moves to the southern edge of the PIC domain ($z=-6$). FTE-B elongates significantly along the dawn-dusk direction. It is twisted at the dawn side so that the axial direction is almost parallel to the z-axis. At the dusk side, FTE-B connects to a newly formed flux rope FTE-C. At $t=660$ s, FTE-B and FTE-C have merged and become indistinguishable. These 3D plots suggest:1) flux ropes arise from multiple X line reconnection and can grow in time along the dawn-dusk direction, 2) the pole-ward moving velocity varies along a flux rope and makes them tilted, and 3) two flux ropes can merge and form a new long rope.\\

\begin{figure}[htb]
\centering
 \includegraphics[width=0.8\textwidth, trim=0cm 0cm 0cm 0cm,clip,angle=0]{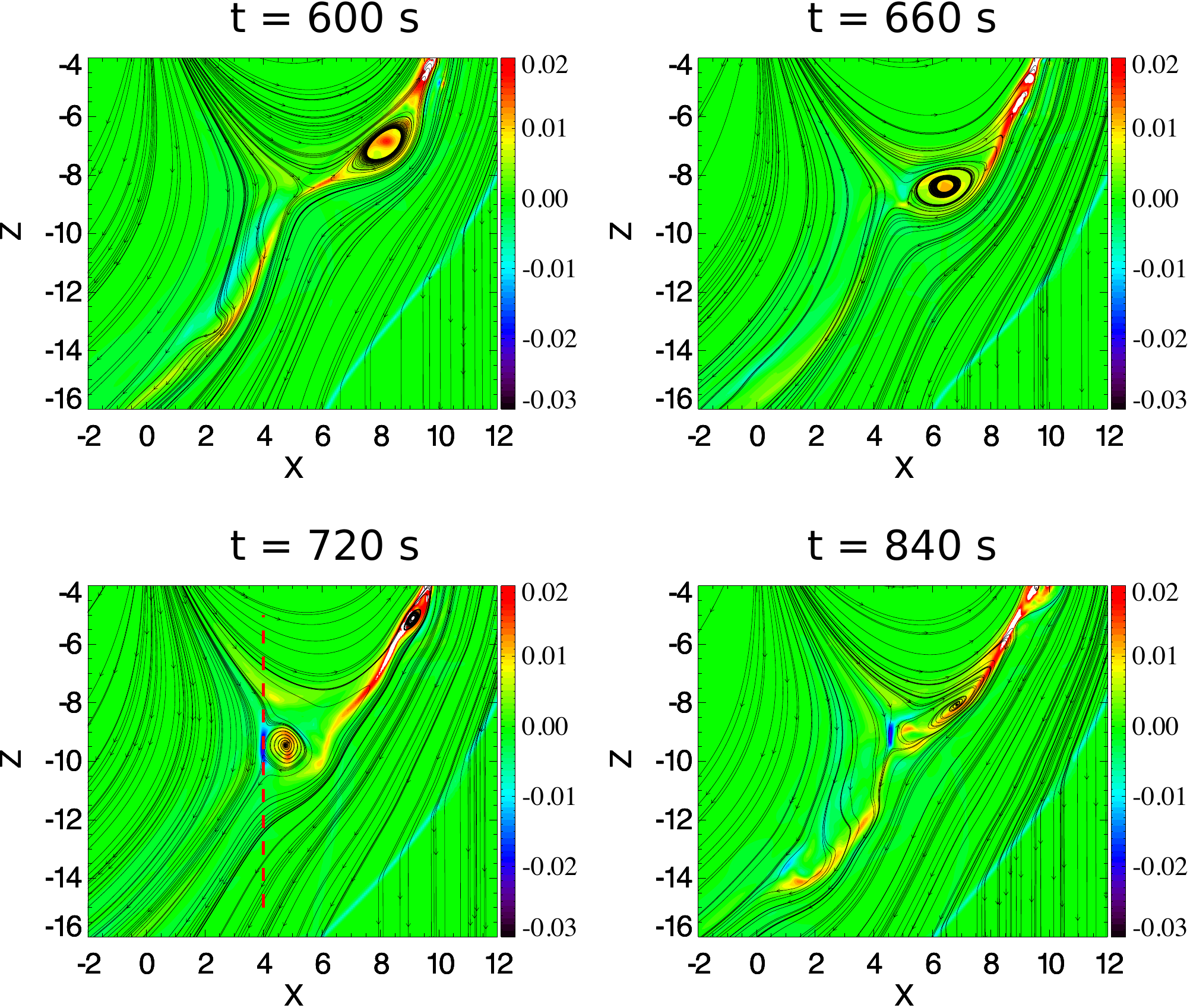}
  \caption{The FTE dissipation when it is crossing the southern cusp. A series of snapshots of current density $j_y [\mu \mathrm{A/m^2}]$ and field lines are shown. The plots are obtained from MHD output. Along the FTE's trajectory, the grid is uniform and the cell size is $1/16\,R_E$. The red dashed line indicates the cut used in Figure~\ref{fig:cusp-1d}}
  \label{fig:fr-dissipation}
\end{figure}

\begin{figure}[htb]
\centering
  \includegraphics[width=0.4\textwidth, trim=0cm 0cm 0cm 0cm,clip,angle=90]{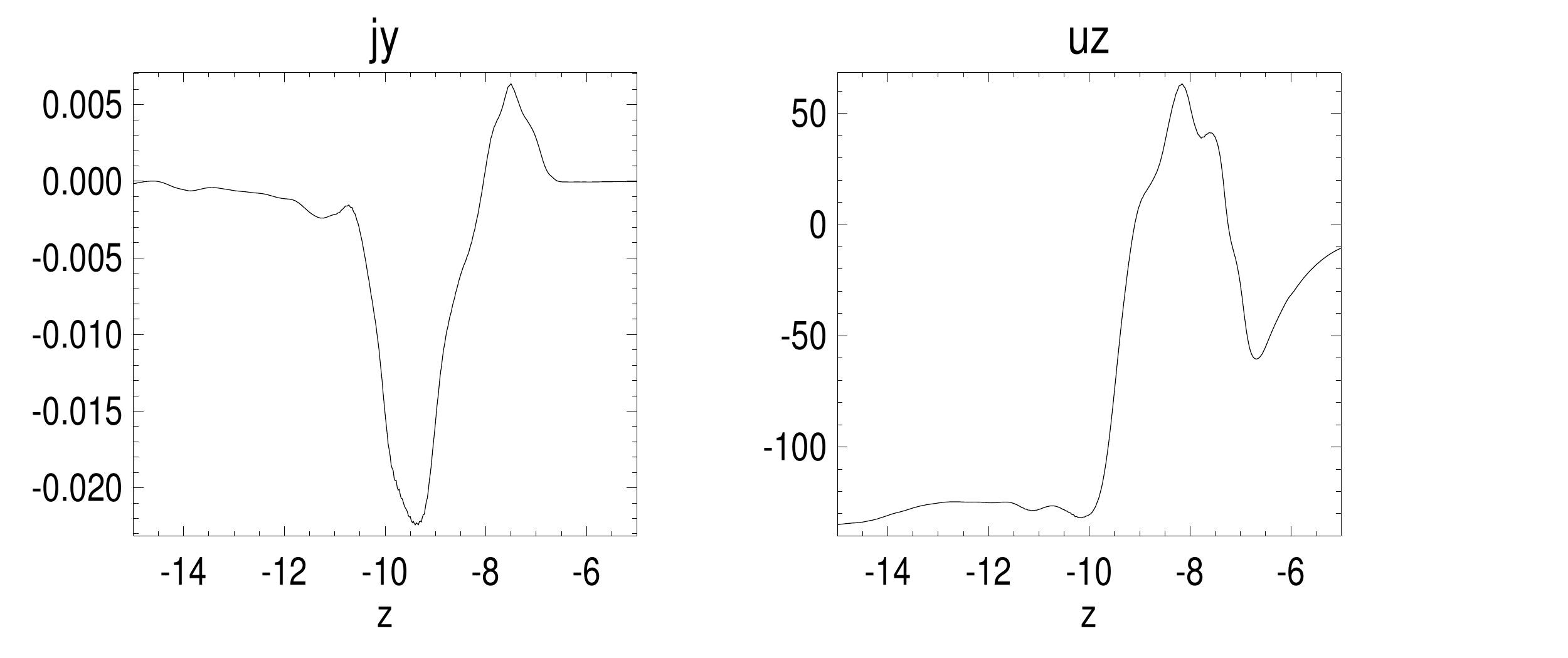} 
  \caption{$j_y [\mu \mathrm{A/m^2}]$ and $u_z [\mathrm{km/s}]$ along the vertical red dashed line marked in Figure~\ref{fig:fr-dissipation}. The jump of $u_z$  around $z\sim -9\,R_E$ implies the occurrence of magnetic reconnection. }
  \label{fig:cusp-1d}
\end{figure}

Since the PIC code is two-way coupled with the MHD model, the flux ropes can smoothly move out of the PIC region. Figure~\ref{fig:fr-dissipation} shows a series of $j_y$ and field lines of FTE-A in the meridional plane after it leaves the PIC domain. FTE-A moves southward along the magnetopause after it is generated near the subsolar point. At $t=600\,\mathrm{s}$, the flux rope is already close to the southern cusp. There is strong axial current $j_y \sim 0.02\mu$A/m$^2$ near the center of the flux rope. As FTE-A moves toward the cusp, $j_y$ inside the flux rope decreases in intensity, which indicates the dissipation of the magnetic helicity, as we can see at $t=660\,\mathrm{s}$. When the FTE reaches the center of the cusp ($t=720\,\mathrm{s}$), the field lines at the leading edge of the FTE and the cusp field lines are anti-parallel and creates a narrow and short current sheet with negative $j_y$ around $x\sim 4\,R_E$ and $z\sim -9.5\,R_E$. The ion velocity $u_{iz}$ at $x=4\,R_E$ in Figure~\ref{fig:cusp-1d} shows a jump around $z=-9.5\,R_E$. The narrow current sheet and the velocity jump imply that reconnection occurs between the flux rope field lines and the cusp field lines. At $t=840\,\mathrm{s}$, after FTE-A leaves the cusp, the signature of the flux rope becomes very weak: even though the magnetic field is still perturbed,  the $j_y$ component is close to zero near the center and no `O' line can be found. Finally, the remnant of the flux rope completely disappears as it moves toward the tail. \textcolor{black}{Here we show the dissipation of FTEs in the meridional plane. But FTEs were observed along the distant tail magnetopause ($x=-67\,R_E$) on the dusk flank \citep{Eastwood:2012}. One possibility to explain the conflict is that these survived FTEs may bypass the cusps and move along the flank from the dayside to tail magnetopause.} \\

\subsection{Magnetic field signature}
\begin{figure}[htb]
\centering
 \includegraphics[width=0.8\textwidth, trim=0cm 0cm 0cm 0cm]{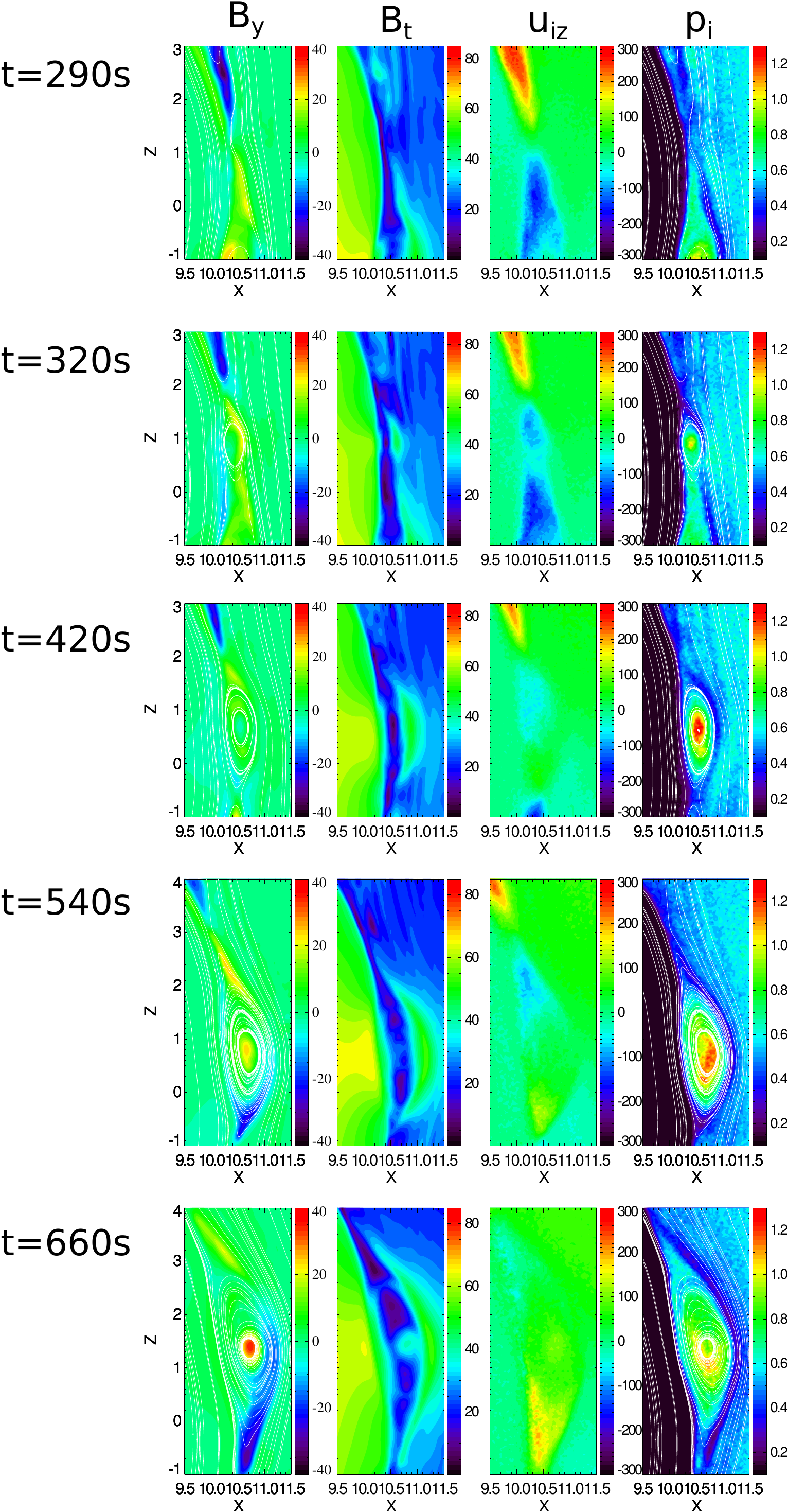}
  \caption{The evolution of FTEs in the meridional plane. From left to right, the four columns show the $B_y[\mathrm{nT}]$ and the projected magnetic field lines; the field strength $B_t [\mathrm{nT}]$; the ion velocity in z direction $U_{iz} [\mathrm{km/s}]$; and the ion pressure $p_i [\mathrm{nPa}]$ overlapped with magnetic field lines.}
  \label{fig:fr-evolution}
\end{figure}

\begin{figure}[htb]
\centering
 \includegraphics[width=0.8\textwidth, trim=0cm 0cm 0cm 0cm]{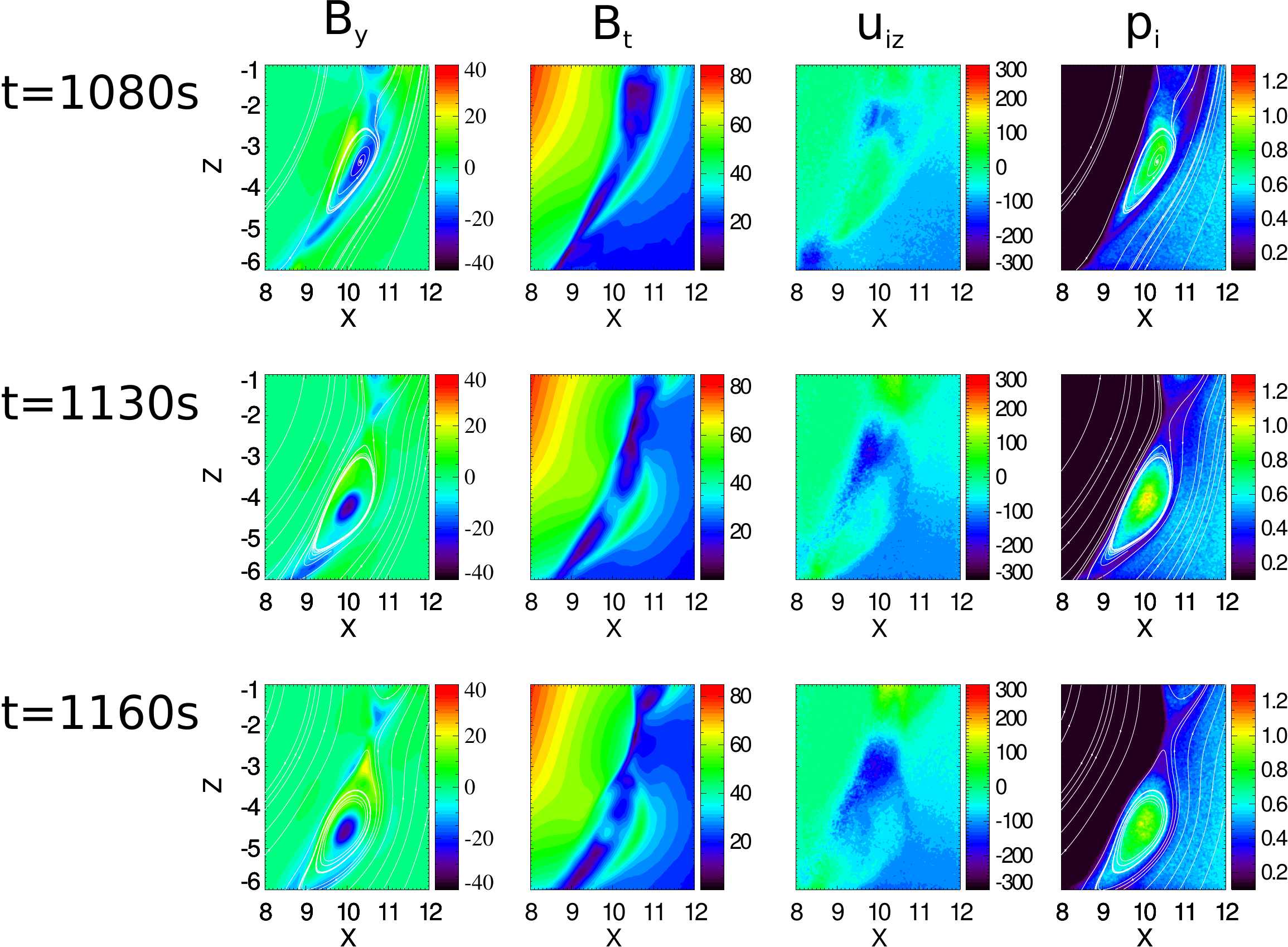}
  \caption{Same as Figure~\ref{fig:fr-evolution}, except that the ion inertial length is scaled up by a factor of 32.}
  \label{fig:f32}
\end{figure}

Since the most widely used indicator of FTEs in satellite data is the magnetic field signature, we discuss how the flux rope magnetic field looks like along a virtual satellite trajectory. A series of meridional cuts are shown in Figure~\ref{fig:fr-evolution} to illustrate the magnetic field evolution. At $t=290\,\mathrm{s}$, north of the FTE-A event, there is an X line at about $z=1\,R_E$ surrounded by the quadrupolar Hall magnetic field $B_y$. As expected, the two branches on the magnetosheath side with amplitude of $\sim 30$ nT are stronger than the other two on the magnetosphere side with amplitude of $\sim 10$ nT. Near the X line, the magnetosheath and magnetosphere are separated by a current sheet accompanied with very weak magnetic field. 30 s later, another X line near $z=0$ arises, and an O line forms between the two X lines. Around the edge of the O lines, the azimuthal component of the magnetic field grows, while the  $B_y$ component is still very weak just near the center. We note that the strong field on the magnetosheath side of the flux rope is mainly contributed by the $B_z$ component because of the accumulation of the inflow magnetic field lines. The reconnection at the northern X line is stronger than the southern one, so the ion jet around the O line is moving southward with a slow speed less than $100\,\mathrm{km/s}$. Inside the O line, the pressure starts increasing. $100\, \mathrm{s}$ later, the pressure at the center of the flux rope already reaches about $1.3\, \mathrm{nPa}$ while the core field is still small. At $t=540\,\mathrm{s}$, the O line structure continues to grow as the two X lines move northward and southward, respectively. We can see the core field $B_y$ at the center of the O line has grown to a significant value of $\sim 30\,\mathrm{nT}$ now, while the center pressure drops to $\sim 1.0\,\mathrm{nPa}$. The converging jets from the two X lines are comparable and the flux rope is almost steady. $180\, \mathrm{s}$ later, the core field grows to $\sim 40\,\mathrm{nT}$ and the corresponding pressure drops to about $0.8\,\mathrm{nPa}$. The whole structure at this stage is moving northward driven by the ion jet generated by the southern X line. \textcolor{black}{To demonstrate the scaling factor has weak influence on the global structures, we perform another simulation with ion inertial length increased by a factor of 32. The simulation results are shown in Figure~\ref{fig:f32}. The FTE in Figure~\ref{fig:f32} shows similar dynamic process as the event in Figure~\ref{fig:fr-evolution}: the core field grows gradually and the ion pressure is anti-correlated with the core field strength. The FTEs in Figure~\ref{fig:f32} and Figure~\ref{fig:fr-evolution} also have comparable sizes.}\\

\begin{figure}[htb]
\centering
 \includegraphics[width=0.8\textwidth, trim=0cm 0cm 0cm 0cm]{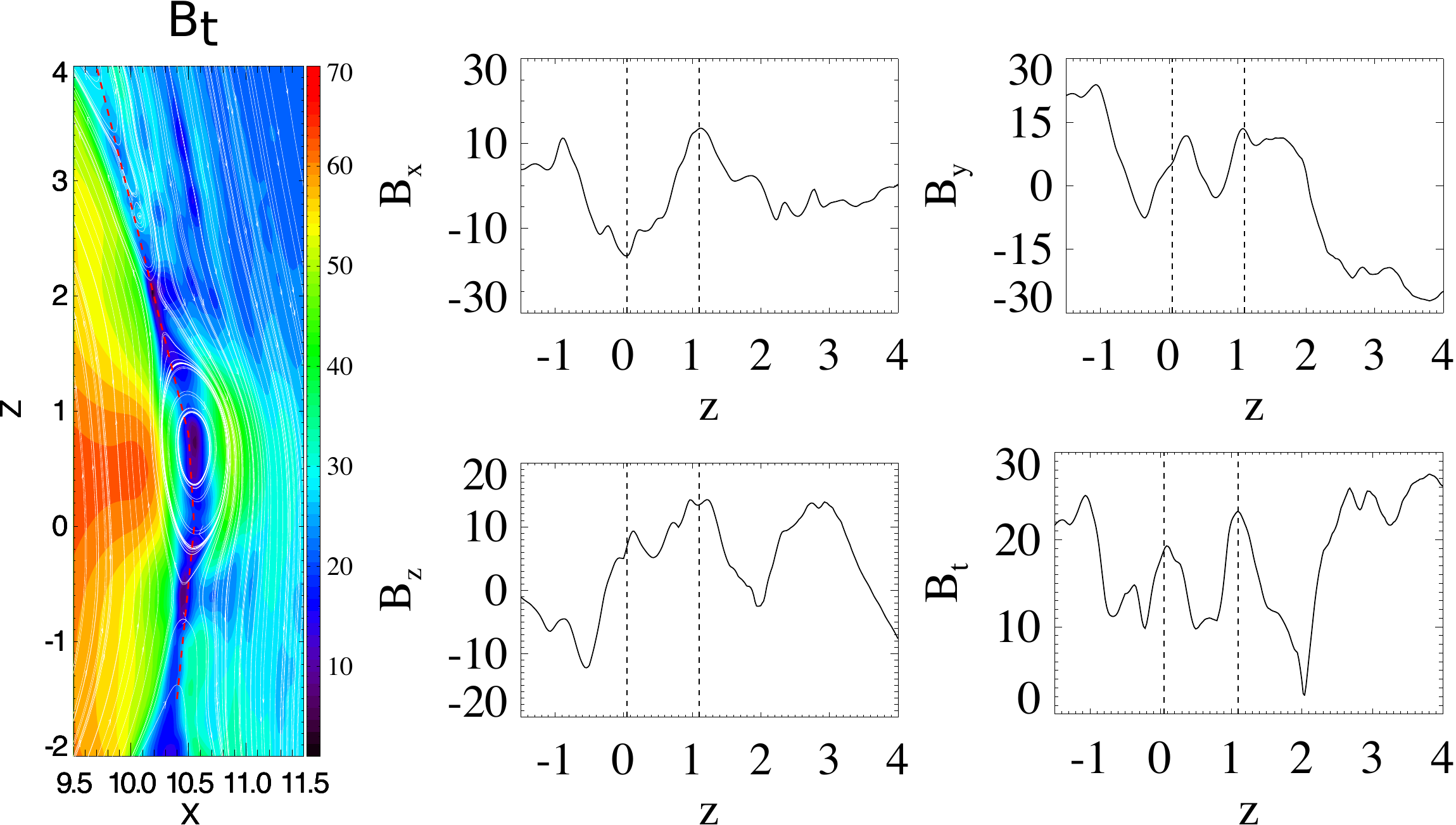}
  \caption{The crater flux rope at $t=420\,\mathrm{s}$. The left panel shows the magnetic field strength and field lines. The right four plots show the magnetic field along the red dashed line in the left panel. The two vertical dashed lines represent the two peaks of $B_x$.}
  \label{fig:c-fte}
\end{figure}

At the early time when the O line just formed, for example, at $t=420\,\mathrm{s}$, the weak core field is surrounded by relatively large toroidal fields. We argue that this is an example of the so-called 'crater FTEs' that have been observed by spacecrafts \citep{Labelle:1987,Zhang:2010}. Since the O line moves slowly during its initial stage of formation, the magnetic field observed at a fixed point can not reflect its global structure. Instead, the magnetic field along the magnetopause (the red curve in the left panel of Figure~\ref{fig:c-fte}) is shown in the right panel of Figure~\ref{fig:c-fte} to illustrate its magnetic field structure. Along the magnetopause, from south to north, the $B_x$ field, which is roughly normal to the magnetopause, reaches a local minimum of $\sim -15$ nT at $z=0$ and then quickly increases to $\sim 15$ nT  at $z=1\,R_E$.  The flux rope is bounded by the depressed magnetic field `trenches' at $z=-0.2\,R_E$ and $z=2\,R_E$ as indicated by $B_t$. The depression results from the low magnetic field strength inside the current sheet as can be seen from the right panel of Figure~\ref{fig:c-fte}. $B_t$ reaches local maximum at the same position of the $B_x$ peaks ($z=0\,R_E$ and $z=1\,R_E$), while the field strength decreases to about $10$ nT between the peaks. We refer to the event on 30 July 2007 observed and analyzed by \citet{Zhang:2010} as a comparison. Figure~6 of \citet{Zhang:2010} shows the magnetic field signature of this event. Even though the 30 July 2007 event has a large guide field (corresponding to $B_y$ component in our simulation), and its magnetic field around the flux rope is more steady than our simulation, the whole structure of this event is similar to what is shown in Figure~\ref{fig:c-fte}.\\

\begin{figure}[htb]
\centering
 \includegraphics[width=0.8\textwidth, trim=0cm 0cm 0cm 0cm]{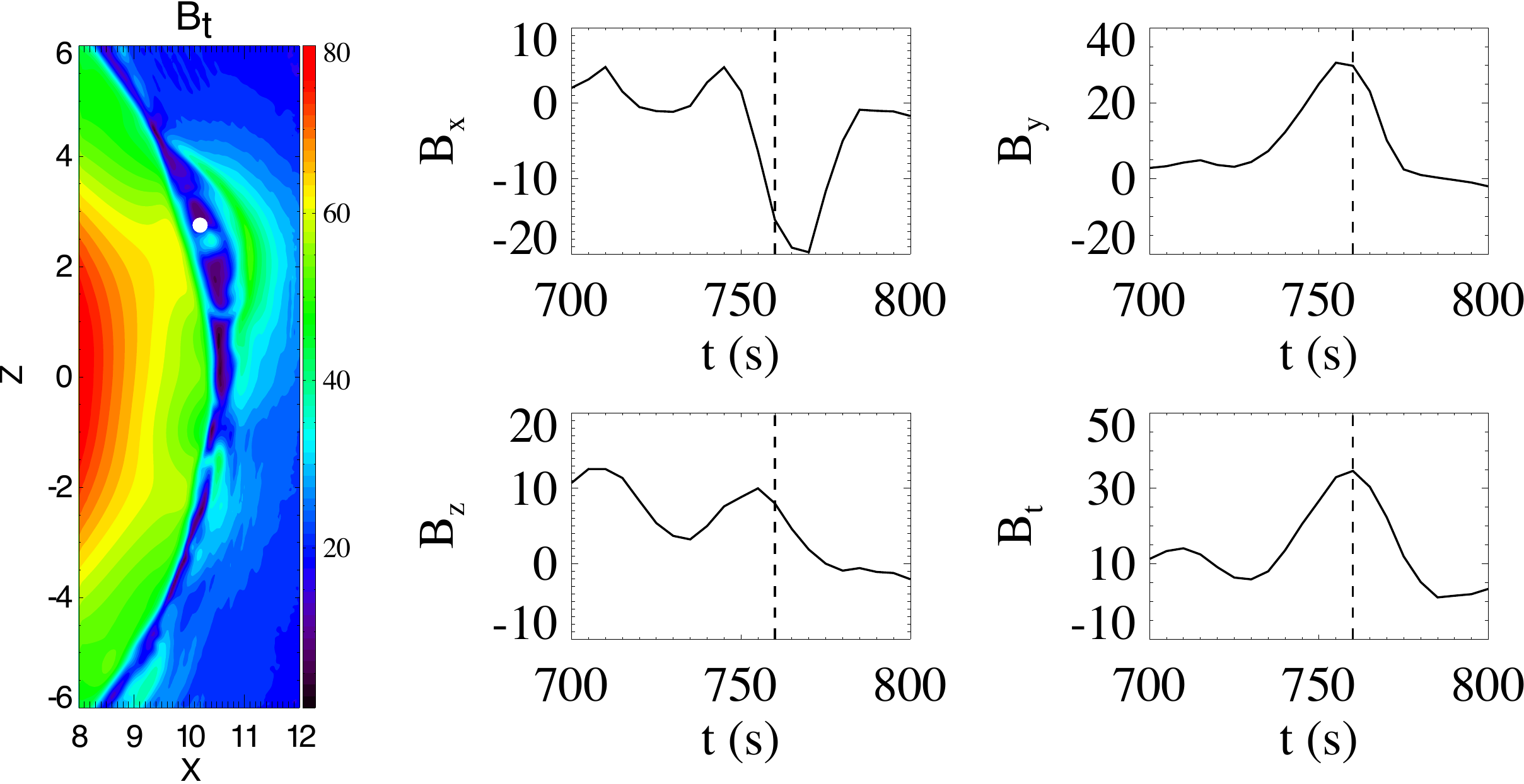}
  \caption{The magnetic field signature of a flux rope with significant core field. The left panel is the magnetic field strength at $t=740\,\mathrm{s}$. The white filled circle at $x=10.2\,R_E$, $z=2.75\,R_E$ is the location of the steady virtual satellite. The right panels show the magnetic field observed by the satellite.  The vertical dashed line at $t=760\,\mathrm{s}$ indicates the location of maximum $B_t$.}
  \label{fig:t-fte}
\end{figure}

As the flux rope evolves, the core field strength grows to a significant value. The magnetic field measured at a fixed position $x=10.2\,R_E$, $z=2.75\,R_E$ is shown in the right panel of Figure~\ref{fig:t-fte}. The vertical dashed line at $t=760\,\mathrm{s}$ represents the location of the maximum $B_t$. Around this time, the $B_x$ field, which is roughly perpendicular to the magnetopause, jumps from $\sim 5$ nT to $\sim -20$ nT within about 25 s. At $t=760\,\mathrm{s}$, both the axial field $B_y$ and the total field $B_t$ reach a maximum. These features match the signatures of an FTE with typical flux rope structure \citep{Zhang:2010}. \textcolor{black}{ During the one-hour long simulation, there are ten FTEs with significant core field moving across the southern PIC edge. The occurrence frequency is consistent with observations \citep{Rijnbeek:1984} and previous MHD simulations \citep{Raeder:2006}.}\\ 

\begin{figure}[htb]
\centering
 \includegraphics[width=0.8\textwidth, trim=0cm 0cm 0cm 0cm]{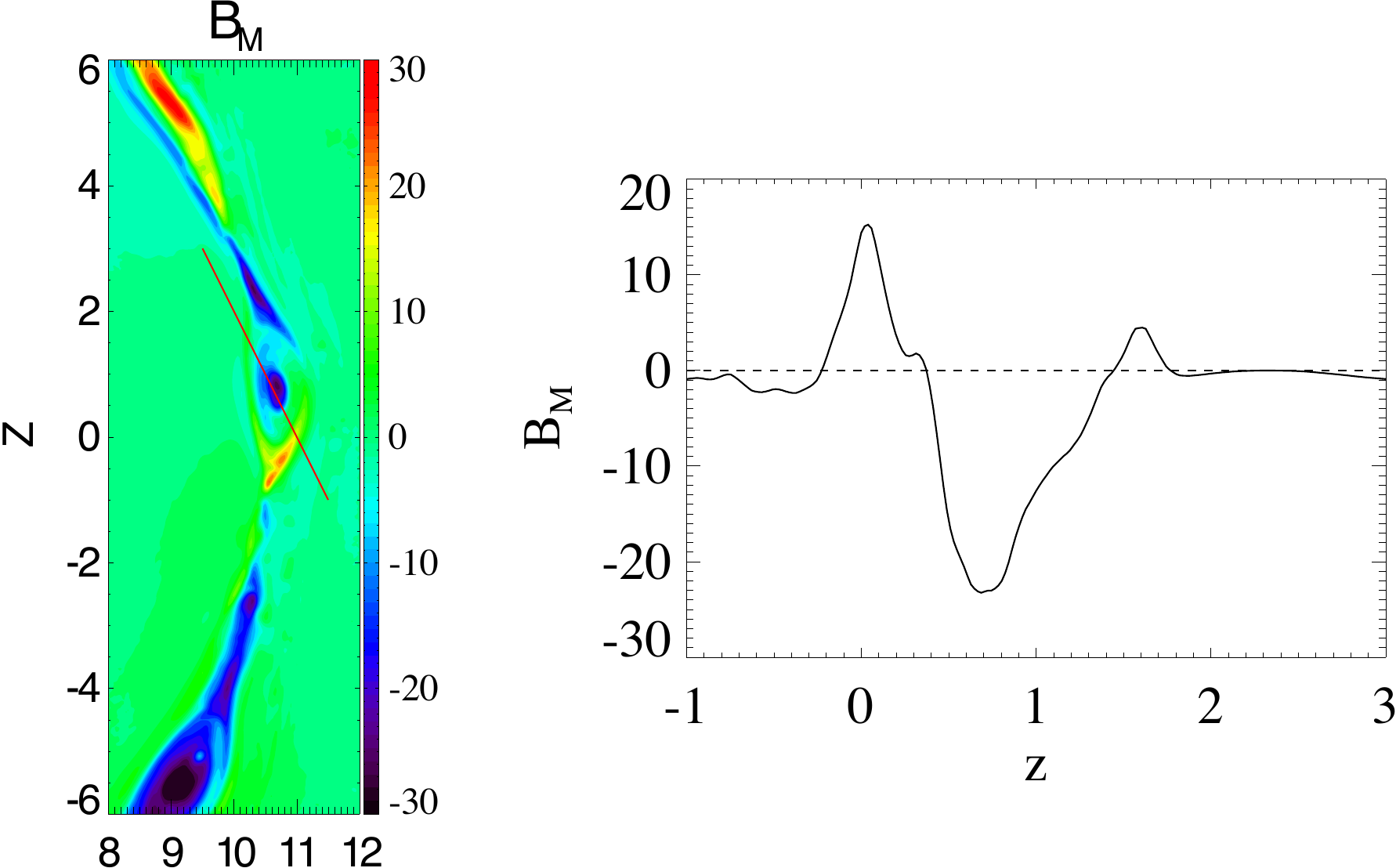}
  \caption{The tripolar guide field structure. The left panel shows the $B_M$ component in the meridional plane at $t=540\,\mathrm{s}$. Around the flux rope center, the guide field is negative, while the southern part of this flux rope is surrounded by the 'Y' shaped positive $B_M$. The field along the red solid line is shown in the right panel.}
  \label{fig:tripolar}
\end{figure}

The IMF is purely southward in our simulation and there is no uniform background guide field at the magnetopause. But a significant core field can still arise during the FTE generation and evolution as seen in Figure~\ref{fig:fr-evolution}. When a flux rope is still close to the X lines, the core field may be encompassed by the Hall magnetic field generated by the reconnection, resulting in complicated guide field structure. The $B_M$ field at $t=540\,\mathrm{s}$ is shown in Figure~\ref{fig:tripolar}.  In order to compare with  observations, the magnetic field has been transformed into a boundary normal coordinate system ($\mathbf{LMN}$), in which the $\mathbf{N}$ component points outward, normal to the magnetopause, the $\mathbf{M}$ component is determined by $\mathbf{N}\times \mathbf{Z}_{GSM}$ and the $\mathbf{L}$ component completes the right-hand coordinate system. Since the plot is shown in the meridional plane, the $\mathbf{Y}_{GSM}$ direction is anti-parallel to the $\mathbf{M}$ direction.  Around the flux rope center, the guide field $B_M$ is negative, while the southern part of this flux rope is surrounded by positive $B_M$. The polarity of the positive 'Y' shape $B_M$ is consistent with the Hall magnetic field generated by the X line at $z=-1\,R_E$. If a satellite is moving across the flux rope along the red solid line in the left panel of Figure~\ref{fig:tripolar}, the satellite will observe a tripolar guide field structure (right panel of Figure~\ref{fig:tripolar}). Similar structure was first observed in the solar wind \citep{Eriksson:2015}, and then it was observed by the Polar satellite at the magnetopause (see Figure~1 of \citet{Eriksson:2016}). The Polar event shows a large negative $B_M$ core field bounded by two narrow $B_M$ depressions in the presence of  a large background guide field. There is no background guide field in our simulation and thus the right panel of Figure~\ref{fig:tripolar} shows a pure tripolar structure: the large negative $B_M$ field is surrounded by two relative small positive peaks. Despite the difference in the background guide field, the topology of $B_M$ obtained from our simulation is very similar to the Polar observation. \\

\subsection{Kinetic features}
\begin{figure}[htb]
\centering
    \includegraphics[width=1.1\textwidth, trim=0cm 0cm 0cm 0cm,angle=0]{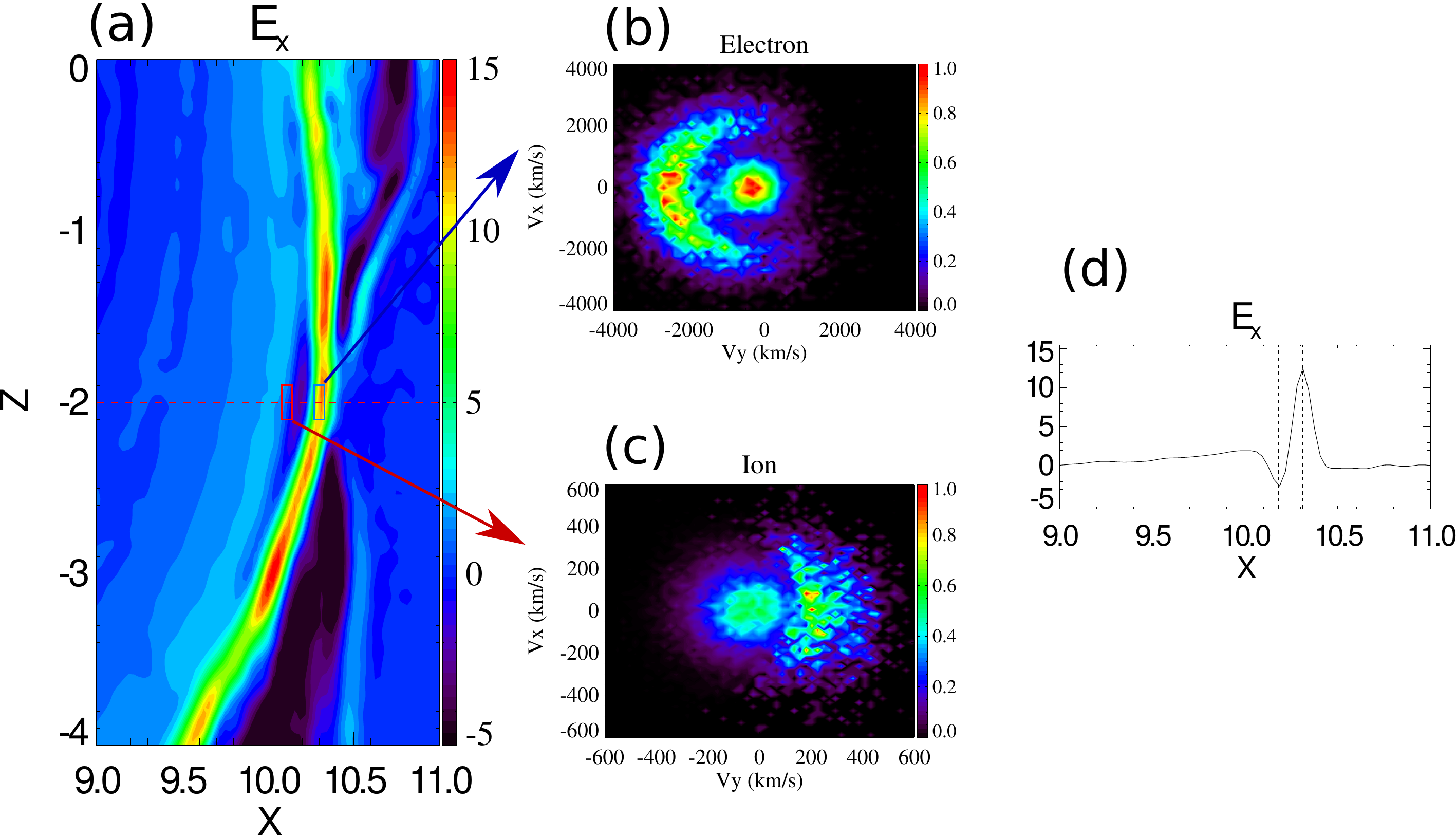}  
    \caption{The Larmor electric field and crescent electron and ion phase space distributions. (a) $E_x [\mathrm{mV/m}]$ in the meridional plane at $t=3600\,\mathrm{s}$.  (b) The normalized electron distribution in $V_y-V_x$ phase space. The electrons are inside the blue box shown in (a): $10.27\,R_E<x<10.33\,R_E, -0.3\,R_E<y<0.3\,R_E, -2.1\,R_E<z<-1.9\,R_E$. (c) Ion phase space distribution for particles inside the red box in (a):$10.08\,R_E<x<10.14\,R_E, -0.3\,R_E<y<0.3\,R_E, -2.1\,R_E<z<-1.9\,R_E$. The phase density is normalized. (d) $E_x$ along the red dashed line in panel (a). }
    \label{fig:crescent}
\end{figure}

We have examined the global structure of the FTEs in the previous discussion.  In this subsection, we will demonstrate that the underlying kinetic physics is properly captured by our model. The Larmor electric field, identified by \citet{Malakit:2013}, is a  localized electric field that appears on the magnetospheric side of the dayside reconnection site. The x-component of the electric field $E_x$ at the end of the simulation (t=3600s) is shown in Figure~\ref{fig:crescent}. The positive $E_x$ pointing towards the Sun along the magnetopause is the Hall electric field, while behind the Hall electric field, the localized negative field pointing towards the Earth is the Larmor electric field. A 1D cut through the reconnection site along the x direction is also shown in Figure~\ref{fig:crescent}. The Larmor field strength is -3 mV/m, the magnetospheric side ambient field is about 2 mV/m, and the nearby Hall field is about 12 mV/m. These values are reasonably close to the MMS observation by \citet{Graham:2016}: the Hall electric field strength is $\sim 20\, \mathrm{mV/m}$ and the Larmor field strength is about 10 mV/m (see Figure~2 of \citet{Graham:2016}).\\

Even though the ion inertial length is scaled up by a factor of 16 in the present simulation, the electric field strength is not sensitive to the scaling factor. Ignoring the electron inertia term, the generalized Ohm's is:
\begin{eqnarray}
\mathbf{E} = -\mathbf{u_i}\times \mathbf{B} + \frac{1}{q_in_i} \mathbf{j}\times \mathbf{B} - \frac{1}{q_in_i}\nabla p_e =  -\mathbf{u_e}\times \mathbf{B} - \frac{1}{q_in_i}\nabla p_e
\end{eqnarray}
\citet{Toth:2017} shows the electron velocity $\mathbf{u_e}$ of the current sheet does not change with the scaling factor while the current sheet width scales. \textcolor{black}{The gradient of electron pressure is inversely proportional to the scaling factor, because the pressure jump is fixed across the current sheet and the current sheet width is proportional to the scaling factor. Since the charge per ion or electron is also reduced by the same factor, the scaling does not change the electric field strength. Besides the scaling of the ion inertial length, a reduced ion-electron mass ratio $m_i/m_e=100$ is used in this study to increase electron kinetic scales (see section \ref{sec:iPIC3D}). The influence of the mass ratio $m_i/m_e$ has been studied in numerous papers \citep{Shay:1998, Hesse:1999, Ricci:2004, Shay:2007, Lapenta:2010}.} For the Larmor electric field , \citet{Malakit:2013} specifically estimates its amplitude to be:
\begin{eqnarray}
E \sim \frac{k_B T_{i}}{q_ir_{i}}
\label{eq:larmor}
\end{eqnarray} 
where $k_B$ is the Boltzmann's constant, $T_i$, $q_i$ and $r_i$ are the temperature, charge per ion and ion Larmor radius of the ions on the magnetospheric side. In the simulation, $q_i$ reduces by a factor of 16 and $r_i$ becomes 16 times larger compared to the realistic situation, while the temperature $T_i$ does not change. So, the scaling of inertial length should not influence the strength of the Larmor electric field. \textcolor{black}{On the magnetosheath side, our simulation shows the ion temperature is about $2\times 10^6 \,K$, magnetic field strength is about 60 nT. Substituting these values into Eq. \ref{eq:larmor} gives $E \sim 5.5 \,\mathrm{nT}$. As mentioned above, the value obtained from simulation is about -3 mV/m.}\\

The crescent shape electron phase space distribution has been observed near the electron diffusion region at the dayside magnetopause by MMS \citep{Burch:2016}. The same distribution is also found in our 3D global simulation. The phase space distribution of electrons inside a cube region: $10.27\,R_E<x<10.33\,R_E$, $ -0.3\,R_E<y<0.3\,R_E$ and $ -2.1\,R_E<z<-1.9\,R_E$ is shown in Figure~\ref{fig:crescent}. The crescent distribution is found in the $V_y-V_x$ plane, corresponding to the two velocity components perpendicular to the magnetic field. The crescent hot electrons are drifting along negative y direction with a speed close to 3000 km/s. The direction of the flow is consistent with the $\mathbf{E}\times \mathbf{B}$ direction, and the velocity of the crescent particles is very close to the MMS observation \citep{Burch:2016}. Slightly further away from the reconnection site, where the Larmor field appears, inside a cube $10.08\,R_E<x<10.14\,R_E$, $-0.3\,R_E<y<0.3\,R_E$ and $-2.1\,R_E<z<-1.9\,R_E$, the ion phase space distribution also presents crescent like shape as it is shown in Figure~\ref{fig:crescent}(c). The crescent ions drift in positive y direction because $E_x$ is negative. We also checked the distributions for particles inside the current sheet but far from the reconnection site, and no crescent distributions are found for either electrons or ions. 

\begin{figure}[htb]
\centering
    \includegraphics[width=1.0\textwidth, trim=0cm 0cm 0cm 0cm]{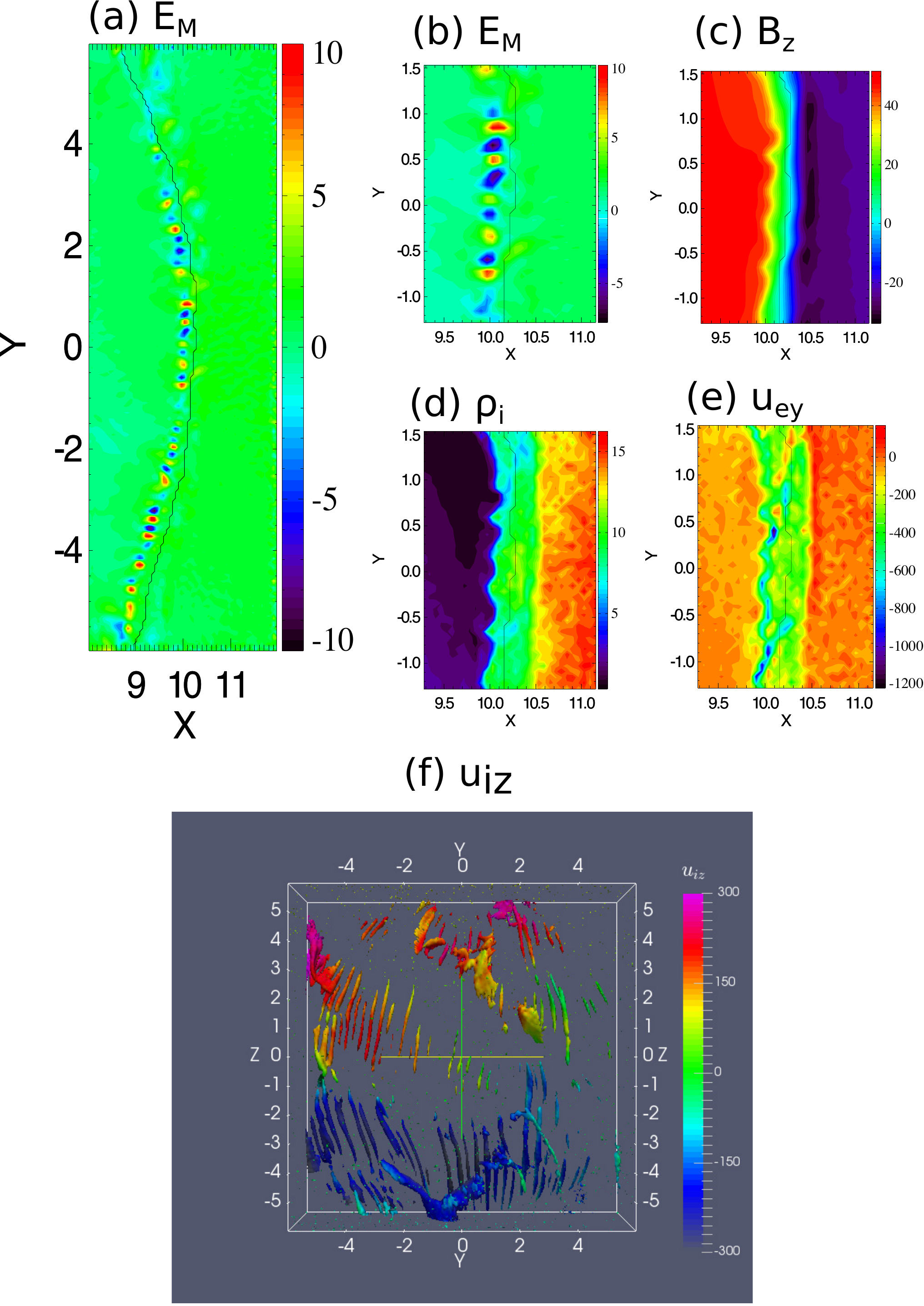}
  \caption{The Low hybrid drift instability (LHDI) at $t=3600\,\mathrm{s}$. (a) Electric field $E_M [\mathrm{mV/m}]$ along the direction that is anti-parallel to the magnetopause current direction in the $z=-3\,R_E$ plane. Near $y=0$, the current direction is almost parallel to the y direction. (b)-(e): zoom-in of different variables for LHDI at $z=-3\,R_E$. (c) is the $B_z$ field in nT, (d) is the ion density in amu/cm$^3$ and (e) is the electron velocity along y direction. The black curves in (a)-(e) separate the negative and positive $B_z$. (f) The 3D contour surface of $E_M=4$ mV/m colored by the ion velocity along the z direction ($u_{iz}[\mathrm{km/s}]$).}
  \label{fig:LHDI}
\end{figure}

Kinetic effects along the magnetopause current direction are also captured by our 3D MHD-EPIC model. Figure~\ref{fig:LHDI} shows the fully developed lower hybrid drift instability (LHDI) at the end of the simulation (t=3600 s) at the $z=-3\,R_E$ plane. The electric field $E_M$ shown in Figure~\ref{fig:LHDI} is the $\mathbf{M}$ component in the boundary normal coordinates, and $\mathbf{M}$ is anti-parallel to the current direction. The black curve in Figure~\ref{fig:LHDI} separates the negative and positive $B_z$. We can see the LHDI appears along the magnetopause on the magnetospheric side. A closer view of $E_M$, as well as $B_z$, ion mass density $\rho_i$ and electron velocity $u_{ey}$ are also shown Figure~\ref{fig:LHDI}. It is clear to see the LHDI arising near the interface of magnetosheath and magnetosphere, where there is a sharp density gradient.  $B_z$, $\rho_i$ and $u_{ey}$ show sawtooth pattern at the same location. The amplitude of the LHDI electric field is about 8 mV/m, which is consistent with MMS observation \citep{Graham:2016}. The dominant wave length shown in Figure~\ref{fig:LHDI}(b) is about $0.38\,R_E$, and the ambient magnetosheath side electron gyroradius is about $r_e=0.025\,R_E$ with the artificially changed charge per electron mass ratio, which results in $kr_e \sim 0.4$, where $r_e = m_e v_e/(q_e B)$ and $v_e$ is defined as $v_e=\sqrt{2T_e/m_e}$. The value of $k r_e$ is also consistent with observation \citep{Graham:2016} and theory \citep{Daughton:2003}. LHDI at different time and different location is analyzed, the value of $k r_e$ varies from $\sim 0.3$ to $\sim 0.5$, and $k r_e \sim 0.4$ is a typical value. Similar as the argument above with the Ohm's law, the electric field strength is not sensitive to the scaling, that is why the LHDI electric field strength agrees with MMS observations. But the length scale does change with the scaling. \textcolor{black}{The charge per mass of electron $q_e/m_e$ is artificially increased by a factor of 294 in the simulation, and the electron thermal velocity reduces by a factor of $\sqrt{18.36}=4.3$ for $m_i/m_e=100$. The magnetic field is realistic, hence the electron gyroradius is about 68 times larger than in reality. If we scale back the LHDI wavelength of the simulation by the same factor, it will be $\sim 35$ km.} As a comparison, MMS observed 10km $\sim 13$km wavelength \citep{Graham:2016}. Figure~\ref{fig:LHDI}(f) shows the isosurfaces of $E_M=4$ mV/m colored by the ion velocity $u_{iz}$ viewed from the Sun. Along the magnetic field direction, the isosurfaces are cut off two or three times. The ion velocity jumps or even change directions across a cut-off region. It suggests these cut-off regions corresponding to the reconnection sites and the LHDI electric field is weak near the diffusion regions \citep{Pritchett:2013}. 

\subsection{Comparison with Hall MHD}
For comparison, we also run a pure Hall MHD simulation with the same setup except the PIC region is removed and the MHD grid resolution around the dayside magnetopause is refined to $1/32\,R_E$, which is the resolution used by PIC in the MHD-EPIC run. Even for Hall MHD, resolving the ion inertial length is necessary in order to capture the Hall effect correctly. Due to the small kinetic scale inside the magnetosheath, scaling the ion inertial length is also required for a global Hall MHD simulation since Hall MHD is also computationally expensive as we will see. We note that the ion inertial length in the pure Hall MHD simulation is also scaled up by a factor of 16 so it can be resolved by $1/32\,R_E$ cell. Hall MHD is reasonably optimized by using semi-implicit scheme to overcome the time step imposed by the whistler mode wave and speed up the simulation. It still takes 6400 cores running about 67 hours to model one hour because of the high resolution and the stiffness of the Hall term. As a comparison, the MHD-EPIC simulation (170 hours on 6400 cores) is about 2.5 times more expensive. Hall MHD produces the Hall magnetic field  near the X line and generates flux ropes in a way similar to MHD-EPIC. But Hall MHD can not reproduce the kinetic features, neither the crescent particle distributions nor the LHDI.\\ 

\section{Summary and conclusion}
We have performed a one-hour long high-resolution global simulation with the MHD-EPIC model to study  dayside reconnection and FTEs. Our simulation is the first attempt to investigate the FTEs and reconnection with kinetic physics resolved in a realistic magnetopause environment. Although the kinetic scale is artificially increased to reduce the computational cost, the model still captures the kinetic features very well. MMS observations, like the crescent particle phase space distribution and LHDI, are reproduced in our model. The FTEs from the simulation also agree well with spacecraft observations. The key results from the present simulation are:
\begin{itemize}
\item When an FTE arises, its cross section is small and it is short in the dawn-dusk direction. During its growth, the cross section increases and the FTE extends along the dawn-dusk direction. 
\item An FTE forms near the subsolar point and moves toward the poles under steady southward IMF conditions. When the FTE reaches the cusp, reconnection happens between the FTE magnetic field and the cusp magnetic field lines, thus dissipating the FTE. The signature of FTE is weak behind the cusps. 
\item FTE is flanked by two reconnection sites during its formation, and the converging ion jets around the FTE are found.
\item The present simulation confirms that the 'crater FTEs' magnetic field signature can be found at the early stage of an FTE formation when the axial magnetic field is still weak. A strong core field may develop as the FTE evolves, and the Hall magnetic field may provides the initial seed core field. Therefore a fully developed FTE has the typical strong core field structure. 
\item A tripolar guide field structure is found from our simulation.
\item The Larmor electric field is found near the reconnection site on the magnetospheric side, and its amplitude is about -3 mV/m.
\item  A crescent electron phase space distribution is found near the reconnection site where the Hall electric field reaches its maximum. A similar distribution is also found for ions at the place where the Larmor electric field appears. 
\item The lower hybrid drift instability (LHDI) appears at the interface of the magnetosheath plasma and magnetosphere plasma. The LHDI electric field peak strength is about 8 mV/m, and a typical ratio between its wavelength and the electron gyroradius is about $kr_e \sim 0.4$. The simulation agrees with the MMS observations and theory. 
\end{itemize}

Compared to the models relying on ad hoc resistivity or numerical resistivity to generate FTEs or investigate reconnection process, our 3D MHD-EPIC model makes one significant step forward by incorporating a self-consistent kinetic description of reconnection into a global MHD model. While the kinetic scales are increased by artificially reducing the charge per mass for both ions and electrons, all the other parameters are realistic. The scaling changes the size of kinetic features, for example the wavelength of LHDI, but other values, like the strength of Larmor electric field or LHDI electric field, are not modified by the scaling. Another  artificial change is the solar wind electron pressure. It is set to a value 8 times larger than the ion pressure so that $p/p_e \sim 2.5$ inside the magnetosheath while the ratio is usually about $4 \sim 12$ from observation \citep{Wang:2012}. The artificially increased electron pressure can help to stabilize the simulation, and it does not deviate significantly from the observed values. We plan to improve this in the future studies. \\

The MHD-EPIC model offers a powerful tool to study magnetospheric physics. The PIC code only covers the dayside magnetopause in the present simulation. As a natural extension, it can be elongated to cover the bow shock so that the kinetic processes associated with the bow shock can be modeled. Another future application is covering the tail reconnection site with another PIC region, so that both dayside and tail reconnections are handled by a kinetic code and we can study substorm in a more realistic way.

%
%
%
%


%
%
%

%
%

%

%

\acknowledgments
This work was supported by the INSPIRE NSF grant PHY-1513379, NSF strategic 
capability grant AGS-1322543, NASA grant NNX16AF75G, NASA grant NNX16AG76G, 
and the Space Hazards Induced near Earth by Large, Dynamic Storms (SHIELDS) 
project DE-AC52-06NA25396, funded by the U.S. Department of 
Energy through the Los Alamos National Laboratory Directed Research and 
Development program. 

Computational resources supporting this work were provided on the 
Blue Waters super computer by the NSF PRAC grant ACI-1640510,
on the Pleiades computer by NASA 
High-End Computing (HEC) Program through the NASA Advanced Supercomputing (NAS)
Division at Ames Research Center,
and from Yellowstone (ark:/85065/d7wd3xhc) provided by NCAR's Computational and
Information Systems Laboratory, sponsored by the National Science Foundation. 

The SWMF code (including BATS-R-US and iPIC3D) is publicly available through the 
csem.engin.umich.edu/tools/swmf web site after registration. 
The output of the simulations presented in this paper can be obtained 
by contacting the first author Yuxi Chen.

%
%
%
%
%
%
%
%
%

\clearpage                                                                                            
\bibliography{csem,yuxichen}                                                                                                  
                                                                                                                              
                                                                                                                              
                                                                                                                              
                                                                                                                              
                                                                                                                              

\end{document}